# Securing Edge Networks with Securebox

Ibbad Hafeez
Department of Computer Science
University of Helsinki
Helsinki, Finland
ibbad.hafeez@helsinki.fi

Aaron Yi Ding
Department of Informatics
Technical University of Munich
Munich, Germany
aaron.ding@tum.de

Sasu Tarkoma
Department of Computer Science
University of Helsinki
Helsinki, Finland
sasu.tarkoma@helsinki.fi

*Abstract*—The number of mobile and IoT devices connected to home and enterprise networks is growing fast. These devices offer new services and experiences for the users; however, they also present new classes of security threats pertaining to data and device safety and user privacy. In this article, we first analyze the potential threats presented by these devices connected to edge networks. We then propose Securebox: a new cloud-driven, low cost Security-as-a-Service solution that applies Software-Defined Networking (SDN) to improve network monitoring, security and management. Securebox enables remote management of networks through a cloud security service (CSS) with minimal user intervention required. To reduce costs and improve the scalability, Securebox is based on virtualized middleboxes provided by CSS. Our proposal differs from the existing solutions by integrating the SDN and cloud into a unified edge security solution, and by offering a collaborative protection mechanism that enables rapid security policy dissemination across all connected networks in mitigating new threats or attacks detected by the system. We have implemented two Securebox prototypes, using a low-cost Raspberry-PI and off-the-shelf fanless PC. Our system evaluation has shown that Securebox can achieve automatic network security and be deployed incrementally to the infrastructure with low management overhead.

*Keywords*-IoT, smart home, security, network, firewall, middlebox, architecture

## I. INTRODUCTION

Recent advancements in technology have been the driving factor in the development of the new generation of smart portable devices including smart phones, smart watches, and tablet PCs to give some examples. Together they add up to more than 3 billion devices and their number is growing at a fast pace [44]. Internet of Things (IoT) has recently gained huge popularity among consumers and estimates predict that more than 20 billion IoT devices will be connected to the Internet by 2020 [7]. IoT devices typically contain sensors and operate on limited computational and power resources. These devices are connected to the Internet either directly or via an IoT hub. IoT devices are primarily used to collect data from surroundings. This data is later analyzed to extract valuable information to be used in different applications.

IoT devices offer many time saving and comfort features for an average user [1]. Users can remotely switch off smart lights or open door locks using their mobile phones. With a number of smart IoT devices launched every day, IoT vows to bring convenience to user's everyday life.

Medium and large scale enterprises are adopting *Bring Your Own Device* (BYOD) policies for allowing their employees and guests to connect their personal devices to the enterprise network. Connecting a large number of heterogeneous devices to the enterprise network has brought a new set of problems for enterprise network security. Since most of the users do not know about malware and exploits potentially installed in their smart devices, any infected device can compromise security of the entire network. Surveys show that majority of Chief Information Security Officers (CISOs) feel that network security operations have become much more difficult to manage compared to the past [6].

Software-Defined Networking (SDN) promises to change the way traditional networks are managed by offering a flexible model that supports innovation [21], [37]. SDN has been used for wide area networks (WAN) and data center environments [34]. SDN has not been applied for the home networks yet; however, we believe that it can support this environment with better security and remote management capabilities. Previous research has showcased techniques for using SDN for dynamic re-routing of traffic through middleboxes deployed outside the network [15], [14], [27].

Security and privacy are important concerns for online users and applications. With the recent popularity of e-commerce, cloud storage and cloud based services, network security and user privacy have become even more important. IoT and BYOD related security threats are fairly new to existing network security techniques and tools, which are mostly designed for large enterprise networks [6]. Therefore, we need to develop new techniques for securing these networks connecting large numbers of heterogeneous devices.

The cost of deploying and operating network security solutions, e.g., firewall (FW), deep packet inspection (DPI) is high. Therefore, these solutions are mainly adopted by large enterprises with sufficient resources to deploy and maintain them. Small enterprise and home users also need similar facilities, but do not have the resources. Our work in this article introduces the advantages of these sophisticated security and remote management solutions to all users with low cost.

The main contributions of this paper are:
- Introducing SDN at the edge for automatic network security, management and bringing the benefits of security services from cloud-based virutalized middleboxes security services to smart home and small/medium enterprise users.
- Proposing *Securebox*: a redesigned, low-cost, remotely manageable home gateway for securing smart home, IoT, BYOD environments and *Cloud-based Security Service*: a cost efficient, scalable security service offering automatic network management, traffic analysis services for detection/mitigation of network threats using collaborative mechanism.
- Implementation and evaluation of *Securebox* and *Cloud-based Security Service* demonstrating feasibility of proposed system in live networks.

**Roadmap:** Section II identifies a set of key security problems in different networked environments. Section III explains the design and architecture of our proposed solution. Section IV describes the implementation details of system prototypes. We evaluate system performance in Section V. We discuss the limitations in current state of the art in Section VI before concluding in Section VII. For simplicity, rest of the paper will refer to IoT, hand-held device as user device and smart home, small enterprise and small office/home office networks as SOHO networks.

## II. BACKGROUND

Typically, the routers or gateways installed in SOHO networks are mainly protecting user devices in the network. These gateways provide Network Address Translation (NAT) features and prevent direct access to the devices from the outside network. However, new generation of IoT devices offer remote management features, which require the devices to expose a management interface to the Internet. Some IoT devices are connected to an IoT hub or user's smart phone which is further connected to the Internet, therefore, providing an indirect way to access these devices.

With more and more devices connected to SOHO networks, they are becoming a lucrative target for criminals. Criminals can remotely break into a user's home network and passively monitor IoT sensor traffic for determining if the user is at home or not, an attacker can remotely open door locks or disable the perimeter security system. This information can be sold and utilized in various ways not approved by the user. Recently, researchers have shown how a connected car can be remotely controlled, which can result in fatal accidents [45].

Most of these devices in BYOD environments are not protected and can be contaminated with malware and spyware. Users also carry their devices to conference rooms and facilities with limited access. The recording instruments, e.g., microphone, camera, GPS can be used to record valuable secret information and transfer it to unwanted entities. Such devices can also infect other devices in the network.

Recently, a number of attacks have surfaced in which millions of devices are hacked to remotely control them for malicious activities. Hundreds of thousands of home routers were remotely controlled as of February 2014 to change their DNS server setting to an attacker's controlled server. These hacked devices were then used to perform phishing, click fraud attacks etc.

Poor device management significantly eases the task of an attacker to remotely access user devices. In late 2014, a hacker searched the Internet for connected CCTV cameras and tried logging in to them using factory default login credentials. The attacker was able to login to thousands of CCTV cameras across the world and obtain live video feed [7]. Similarly, there have been incidents where personal computers were hacked to record live footage from webcams and used for blackmailing and extortion [61].

### A. Motivation and Problem Statement

With the continuous evolution and growing trend of security and privacy attacks using personal computing and smart devices, there is a need for improving network security by monitoring and auditing device activity and detecting security issues [5]. SOHO and smart enterprise networks are either poorly managed or not managed at all due to lack of resources, vigilance and motivation on behalf of users. As a result, an attacker can break into these networks to gain access and potentially control the devices inside these networks. These compromised devices can then be used for spying on user activities, click-fraud, phishing, Distributed Denial of Services (DDoS) attacks, bitcoin mining etc. These issues put SOHO networks at the center of network security picture [10].

Users would like to have network devices that are secure and easy to manage [20]. Recent research has shown that users are more comfortable and enthusiastic to manage their networks when they obtain more information about their network activity [20], [19]. The need for making network management easier and simpler is more prominent in SOHO networks, because it is neither feasible nor scalable to hire experts who can individually manage the security of each network [18].

In our research, we conducted a user study of 150 users from academia and industry to assess user perceptions in home network security management. Our user study of 150 users has revealed that majority ($\geq 80\%$) of users find it very difficult to manage their network access points and gateways and require an easy to operate version of these devices. Therefore, average users should not be burdened with the complex tasks of managing home network. Our work envisions a network gateway which offload management and operational tasks to an external entity (i.e., network management service provider), which has more resources and expertise to perform these operations.

Our proposed gateway will act as a sensor in the network and collect traffic statistics and insights to share them with a service provider. Service provider will use these statistics to (re)configure all gateways in real-time in providing better security against attacks and malicious activities. This model helps in managing networks more efficiently, because the service provider will have a better view across multiple networks and will be able to make well-informed decisions. The broad view of network will also help to identify suspicious traffic trends which might have gone previously unnoticed .

*B. IoT threats analysis*

IoT environments introduce many heterogeneous devices running a variety of protocols and software versions. Due to their small size, low power and limited resources, many of these devices are not even running an operating system. These devices are mainly developed by startups or fast moving teams in enterprises, which work on limited budget and resources. These teams are hurried to develop and launch their products to the market. Therefore, security is often neglected during the design and development of these products. There is usually a lengthy, if any, update cycle for most of these devices. Owing to the number of sensors on these devices with no software updates, a number of security threats are raised against these devices.

In enterprise BYOD environments, presence of heterogeneous devices make the issue more grave for network security team. Table I presents an overview of design and limitations in current state of the art for privacy and security in home networks.

Due to resource constraint, it is very hard to implement security features on IoT devices. Some manufacturers use Trusted platform hardware (TPM) and hardware based scheme for securing these devices [46]. However, this approach is not feasible because of the limited computational and power resources available on IoT devices. A smart solution is to provide fully authenticated and verified access to data collection and operations of these devices, so that no attacker can hijack a device to steal the data or spy on user [2]. Since a majority of malware spread among devices, we should also inspect and restrict uncontrolled device to device (D2D) communications for presenting the devices from infecting each other. In addition, several IoT hubs available on the market lack of such security features [47], [48]. Our proposed cloud-assisted gateway is designed to provide these features for IoT and BYOD environments.

III. CLOUD ASSISTED SECUREBOX

Based on the set of issues identified in the previous sections, this article proposes a new architecture for remote network management in smart home and small enterprise environments.

*A. Overview*

The proposed solution consists of two key components, i.e., *Securebox* and *Cloud-based Security Service* (CSS). The proposed system is designed to scale in different networks including smart home, SOHO and small/medium/enterprise (SME) environments with multiple offices.

The client end, i.e., Securebox uses SDN for network management and operations. Securebox can provide features e.g. device isolation, authenticated device to device (D2D) communication, identification of infected/compromised devices, security profiling of devices etc. whereas CSS can provide cost efficient and scalable detailed traffic analysis and network management features.

*B. Client Edge*

Securebox is a modified gateway running SDN controller and OpenVswitch. Figure 1 shows the architecture of Securebox. Securebox has a local policy database (Pol-DB) which contains security policies for different traffic classes along with specified actions. Securebox is a cheaper replacement for contemporary manageable home gateways in user networks. It provides wired and wireless interfaces to connect user devices. All the traffic flowing to/from the network passes through the Securebox which enforces network policies on this traffic. Securebox delegates all the traffic analysis, security and network management tasks to the CSS.

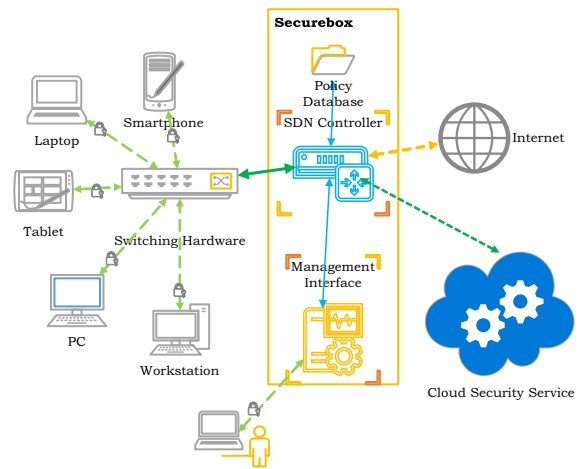

Figure 1: Securebox architecture

*1) Deployment:* Securebox comes pre-configured to connect to the service provider specified CSS and user only needs to connect it to Internet. Securebox management and Pol-DB update are automatically handled by the CSS. Therefore, Therefore, users are free from manually configuring and managing their gateways. Securebox provides an interface showing stats about bandwidth usage per device, security risks detected and removed, suspicious traffic to/from devices in the network, D2D communications etc.

| Current State of Art | | |
|---|---|---|
| **Research** | **Contribution** | **Limitations** |
| A. Brown et al. [19] | **HomeNetViewer**: tool for collecting, annotating domestic network NetFlow records. | Neither addresses security, privacy challenges in home networks nor IoT specific issues. |
| M. Chetty et al. [20] | **uCAP**: tool for monitoring network bandwidth usage in home networks. | Does not address security and privacy challenges for IoT, other devices in home network. Does not audit network traffic to detect suspicious activities in network. |
| A. Alwabel et al. [22] | **SENSS**: an interface for querying ISP to detect anomalies. | Needs to modify ISPs. Does not address challenges from IoT, traffic analysis. Requires expertise to make use of queried information. |
| R. Meyran [23] | **DefenseFlow**: SDN application that programs networks for DoS/DDoS security. | Aimed specifically at enterprise network, high cost, does not address IoT challenges, (currently) limited to DoS/DDoS detection only. |
| J. Sherry et al. [24] | Deploying **middleboxes** in cloud for scalability and cost efficiency. | Addresses only large enterprise use-case, does not talk about security or privacy challenges in IoT domain or SME networks. |
| Y. de Montjoye et al. [3] | **OpenPDS**: Personal metadata management framework allowing user to collect store and manage third party access to their metadata. | Needs support by software products and services for deployment, needs effort from user to manage storage and access of data to third parties. No design support for feedback to the user. |
| H. Haddadi et al. [30] | **Databox**: collects personal data for user and providers controlled access to this data for third parties. | Needs realization of concept to access real world applicability. Security services should be redesigned to support databox. User needs to manage the controlled access to this data. |

Table I: Current state of the art: Contributions and limitations

After configuring CSS, a user profile is set up at CSS for traffic analysis tasks. Securebox then receives a Pol-DB update from CSS. This update consists of a basic set of network policies and is stored in Pol-DB. Later, Pol-DB is regularly updated by CSS, see Section III-C3.

Figure 2 shows the deployment architectures for Securebox where devices are directly connected to the Securebox running an SDN controller and OVS, see Fig. 2a and devices directly connect to it. Scenario B, shown in Fig. 2b, shows the Securebox running an SDN controller and managing OF-capable switches and wireless access points (AP) to which user devices are connected. Securebox is also connected to the CSS and Internet as expected.

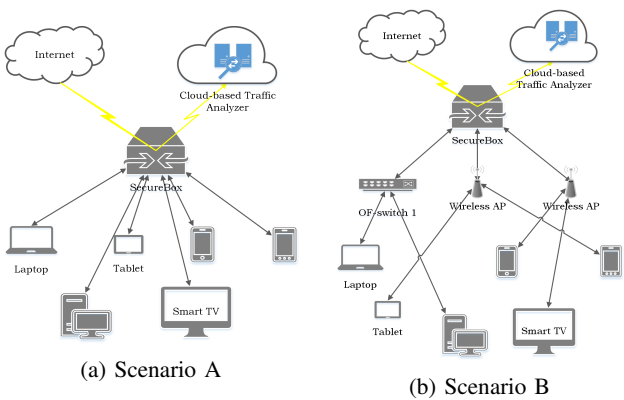

(a) Scenario A      (b) Scenario B

Figure 2: Securebox deployment

Securebox mainly gets the network policies from the CSS. However, Securebox design offers flexibility for power users to configure policies of their own choice e.g. A user can configure that "*Allow my CCTV to connect to my file server (for video feed storage)*". These manually configured policies can be devices specific and have a higher priority than those received from CSS. Securebox can also provide security ranking for each device connected to the network, based on device activity, giving user a better understanding device behaviour. It can generate warnings for the user whenever a suspicious activity is detected and quarantined. These warnings can be displayed on web interface or delivered to user's smart phone.

*2) Functioning:* The flow processing algorithm of Securebox is presented in Algorithm 1. When a new traffic flow is initiated to/ from a device in the local network, Securebox interrupts this flow, extracts some information (6 tuple) from this flow and checks Pol-DB for any matching policy for this traffic flow. If a matching policy is found, the associated decision to allow or drop this traffic is applied to the traffic flow.

If there is no matching policy available, Securebox sends this data to the CSS, which analyzes this data using the user preferred mechanism and returns a security policy. Securebox stores this policy in Pol-DB and applies the decision to the requested traffic flow. In future, similar traffic flow request will get matching policies from the Pol-DB and Securebox will not need to send the traffic to CSS for analysis.

If user installs a CCTV camera for a perimeter security system and the CCTV camera opens a connection to an arbitrary server on the Internet to send video feed, Securebox will intercept this new traffic flow request. Securebox will then extract and send some statistics to CSS for analysis which will identify that user's CCTV camera should not be allowed to connect to any arbitrary server. It will formulate a security policy directing the Securebox to deny any traffic flows which tries to connect CCTV camera to any arbitrary server outside the network. Securebox will implement this policy and deny connection request from the CCTV camera. Any subsequent requests from the CCTV camera to connect to the same or any other arbitrary server will be denied.

Securebox needs to have connectivity with the CSS in order to get the traffic analysis and Pol-DB updates in real time. However, system design allows Securebox to work even when connection to CSS is not available. In that case, Securebox uses Pol-DB to make decisions for incoming traffic flow requests and implements an implicit allow/deny decision rule for traffic flows with no matching policy available.

Another solution is to push some decision-making to the Securebox. This solution will increase hardware requirements for the Securebox therefore increasing the costs and complexity of system. This solution will require more user interaction and most users would not be comfortable with managing all these services.

Sureboxes also act as sensors in the networks to collect information about the network activity and report it back to CSS. CSS then analyzes this information to make better informed decisions for enhanced network and security management and improve quality of service (QoS) to enhance user experience.

*C. Cloud-based Security Service*

CSS is a low cost, highly scalable, service-based solution running in the cloud environment to provide security services including traffic analysis through middleboxes, malware and botnet detection to the clients. It allows any subscriber to run personalized traffic analysis services in the cloud environment at reduced costs and improved scalability.

*1) Architecture:* Figure 3 shows the architecture of CSS. *Cloud Manager* is the central component of the whole system. Cloud Manager is responsible for handling client requests, managing resources, deploying and maintaining middleboxes, handling traffic analysis tasks. Cloud manager delegates some of the sub-tasks to other entities in the CSS.

The *certification authority* manages the certificates for the system and all the subscribers, i.e., Sureboxes. The impact of certification authority on system security is explained in detail in Section V-F. CSS also runs "*Backup Cloud Manager*" which is a state-aware, hot-swappable replica of Cloud Manager which can replace the cloud manager in case it goes down. This improves fault-tolerance and scalability of the system.

**Algorithm 1** Securebox flow processing algorithm
connect to CSS
bootstrap policy-DB
**while** *traffic_flow_request* **do**
　# Extract metadata from incoming connection requests
　$metadata \leftarrow extractMetadata(traffic\_flow)$
　# if matching policy exists in policy database
　**if** *policy_exists(metadata)* **then**
　　# extract decision from matching policy
　　$policy\_decision \leftarrow getDecision(metadata)$
　　# insert traffic flow and update log
　　$insertFlow(OF\_switch, traffic\_flow\_request)$
　　$updateLog(event)$
　**else**
　　# get decision from cloud-security-service
　　$policy \leftarrow getSecurityPolicy(metadata)$
　　# insert traffic flow cache security policy and log event
　　$insertFlow(OF\_switch, traffic\_flow\_request)$
　　$updatePolicyDB(policy)$
　　$updateLog(event)$
　**end**
**end**

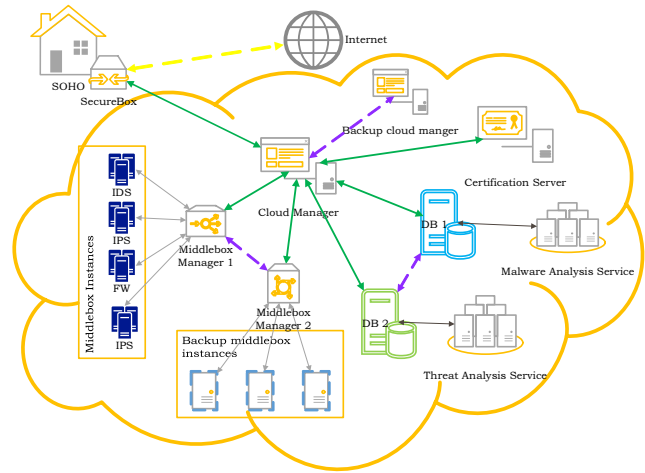

Figure 3: Cloud-based security service architecture

Several kinds of middleboxes, e.g., Intrusion Detection System (IDS), Intrusion Prevention System (IPS), FW can be deployed in the cloud environment. *Middlebox Manager* manages the deployment and operations (e.g. load balancing, fault tolerance) of middleboxes deployed for traffic analysis. Some of the middleboxes are analyzing traffic from delay sensitive applications or enterprises so CSS maintains hot-swappable state-aware replicas of theses middleboxes to improve fault tolerance and efficiency of the system. The instances of middleboxes running in the backup middleboxes pool can immediately swap any middlebox which fails during operations.

There is a huge volume of network traffic flowing to CSS from multiple subscriber networks. This enormous amount of data can be utilized to extract valuable statistics about network traffic, QoS, devices connected to the network. CSS runs various kind of anlaysis on this traffic to extract valuable insights from the traffic. Figure 3 shows the threat and malware analysis services running in CSS. These services use traffic statistics collected from Secureboxes to detect malwares, botnets and other malicious traffic flowing through various networks. The broad view across a number of networks helps in detecting the tiny traces of malicious traffic which usually goes undetected through the traditional network perimeter security systems.

*2) CSS Functioning:* After user profile is created for CSS, all subsequent traffic analysis requests from the (user specific) Securebox are handled according to the preferences of user profile. When a traffic analysis request arrives from a registered Securebox, request handler either sends it to dedicated middleboxes or traffic analysis service depending upon the user profile. The traffic analysis service returns a decision for the incoming request and request handler sends this decision to the Securebox in form of a security policy. Algorithm 2 presents the different operations pertaining to request processing.

Following the example given in Section III-B when the Securebox requests a decision on whether to allow CCTV camera to connect to arbitrary server, CSS analyses the incoming information. CSS maintains a list of addresses of safe/ known servers and the requested connection to unknown server will be flagged suspicious. CSS will generate a policy to "*drop any traffic from CCTV which does not go to (specific) well known servers on the Internet*" and send it to the Securebox.

Enterprise subscribers can use leased middleboxes for analyzing their traffic (as it offers high availability and low processing delay). On the other hand, subscribers from SOHO networks can analyze their traffic in the middleboxes run by CSS for general traffic analysis. Every subscriber can configure the type and sequence of middleboxes used for analyzing their traffic, see Section III-C4. Subscribers can opt-out to share their traffic statistics to be used in other analysis services, as explained in Section V-C

CSS can share the traffic analysis results with other service providers to help them improve their services, QoS and user experience. It can also provide interfaces to third parties for running several kind of analysis on network traffic statistics collected by CSS, for detecting botnets, malware and track suspicious servers hosted over the Internet. This massive and diverse collection of network level traffic statistics can also be very useful for research community as well.

**Algorithm 2** CSS request processing algorithm
---
bootstrap system, services
launch middleboxes
**while** *incoming_analysis_requests* **do**
    # extract information from incoming request
    $info \leftarrow extractInfo(incoming\_request)$
    # if matching policy exists in security-policy-store
    **if** *policy_exists(incoming_request)* **then**
        # get security policy from store and send to client
        $sec\_policy \leftarrow getPolicy(incoming\_request)$
        $sendToClient(sec\_policy, incoming\_request\_id)$
        # update logs
        $updateLog(event)$
    **else**
        # Extract user profile and perform required security analysis
        $user\_profile \leftarrow getUserProfile(incoming\_request)$
        $sec\_policy \leftarrow analyzeRequest(user\_profile, incoming\_request)$
        $sendToUser(sec\_policy, incoming\_request\_id)$
        # cache security policy (if allowed by agreement) and update logs
        $storePolicy(sec\_policy)$
        $updateLog(event)$
    **end**
**end**

*3) Policy-DB updates:* The decisions made for the incoming traffic analysis requests are cached in the *Policy Store* managed by CSS. The information obtained from the threat analysis services run by CSS is also aggregated and stored in the form of network policies. CSS generates regular updates for Secureboxes from the policies collected in Policy Store. These updates are issued to all connected Secureboxes to improve their ability to handle more traffic locally and immediately block any attempts to attack the network. These policy updates from CSS enable smooth functioning of Securebox without requiring user participation.

In the CCTV camera example, when a decision is made that CCTV camera should not be allowed to connect to any arbitrary server (other than well known servers), the next policy update will transfer this security policy to all connected Secureboxes. The next time if a CCTV camera connected through any of these Secureboxes will attempt to connect to any arbitrary server, the connection will be refused by the Securebox without sending an analysis request to the CSS.

The policy update mechanism provides a number of advantages. It significantly reduces the number of traffic analysis requests sent to CSS which reduces the burden on the service provider infrastructure as well as reduces the uplink traffic and saving precious bandwidth. A number of these benefits are explained in detail in Section V-B.

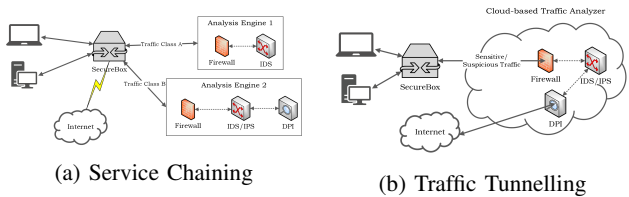

(a) Service Chaining     (b) Traffic Tunnelling

Figure 4: Service chaining

CSS maintains a complete history of all previous updates and current policies in the policy store. At every update, it refines and includes only those policies which were not previously sent. There is a trade-off for the frequency of issuing these updates. If updates are issued too frequently, the size of each update will be smaller, update cycle will be faster and security attacks and threats will be detected more quickly across all networks. However it will also result in more traffic to the Securebox, hence consuming more bandwidth of the user. On the other hand, less frequent updates will use less bandwidth and will increase delay in dissemination of policies required to block network attacks.

In the proposed system design, high priority policies are immediately updated to the Secureboxes whereas policies having less priority are bundled together and sent to the user during hours of lesser network activity e.g. nighttime.

*4) Service Chaining:* The CSS architecture provides support for service chaining for the subscribers to easily combine various kind of network services for analyzing their network traffic. Fig. 4a shows the case where user has chosen separate traffic analysis techniques for different classes of traffic being analyzed e.g. User A has configured that all traffic from IoT devices (i.e. smart fridge, smart TV) belong to class A and traffic from smart phone, tablet or personal computing devices should be classified as Class B traffic. CSS will analyze traffic from each class through a separate set of middleboxes chosen by the subscriber. Traffic from Class A will be analyzed by a FW whereas Class B traffic will be analyzed by FW, IDS and DPI instance running in the CSS environment.

This feature allows subscribers to save the cost and run multiple kind of analysis on its traffic classified into various classes. The proposed architecture allows the user to dynamically modify the analysis services chained together, providing complete control over the traffic analysis being performed on user traffic.

Figure 4b shows the scenario where the user has configured the Securebox to route all the traffic through the CSS. This features allows user to perform middlebox analysis on the whole traffic session involving suspicious traffic. This scenario is especially useful for enterprise users who would like to have all the traffic from guest devices to pass through a set of middleboxes. Subscriber can add/ remove middleboxes on the path of this traffic and using cloud resources offers better scalability as the user traffic volume increases, hence preventing middlebox deployment to become a bottleneck and degrade user experience. Similar concepts have been introduced for re-routing the traffic through middleboxes deployed elsewhere and their work can be used as a feature in the proposed system [27], [24].

*D. User subscription models*

The proposed architecture allows users to reserve a single or set of dedicated middlebox instances for their traffic analysis, achieving low latency and high availability. It also allows to subscribe for traffic analysis services without leasing any middlebox instances. The latter solution will offer lower costs and slightly higher latencies. Section III-C4 explains how a subscriber can benefit from service chaining features offered by the CSS.

Another subscription model allows user to receive periodic network policy updates from CSS without actively analyzing their traffic in middleboxes. All these subscription models include remote management and updates for the Securebox as well.

*E. Deployment Models*

The proposed architecture offers three deployment models for CSS examined below.

*1) Third party security service provider:* In this model, user gets a pre-configured Securebox from third party service provider. Power users can configure Securebox to connect to the CSS of their choice. Service provider can run different kind of traffic analysis on subscriber traffic and get monetary benefits from the traffic statistics collected from various connected networks. This data is valuable for the IoT device manufacturers, service providers (e.g. Netflix [1])., Internet Service Providers (ISPs). It can lead to development of new technologies with built-in security features offering better user experience and QoS.

*2) ISP based deployment:* ISPs can also deploy CSS to provide network management services to their customers. In typical deployments, ISPs provide a home gateway which can be modified to work as a Securebox. ISP's adoption of proposed system will be useful for both customers and ISP. Following this model, customers would not need to install a new gateway and ISP can get valuable information about the user networks to offer distinguished and personalized services. This model will save the cost of deployment and operation for both customers and ISPs and improve ISP operations.

*3) Private deployment:* Private deployment model is useful for research and enterprise-scale deployments since it provides a complete control over the infrastructure. In this model, a client (e.g. enterprise) deploys its own CSS and the Secureboxes are managed by using this private CSS. This model provides a central control interface to monitor and operate network across all deployments. All the traffic is analyzed in a centrally managed infrastructure where personalized traffic analysis techniques can be applied to the data. Private deployment model reduces any privacy concerns since the network information is not shared to third party to any external entity.

---
[1]www.netflix.com

In traditional networks, it is possible to deploy middleboxes centrally at a gateway location and traffic from different establishments is routed through the centrally deployed middleboxes. However, these gateways frequently become bottlenecks, resulting in bad user experience. This model offers very little flexibility for configuration, management and operations in live deployment. However, the proposed system will offer more flexibility by enabling network managers to classify the traffic and change the middleboxes on the fly. It will greatly improve fault tolerance by significantly reducing downtime of middleboxes. It also improves scalability of infrastructure during peak access periods without compromising user experience and network operations.

## IV. System Implementation

We have implemented two prototypes for evaluating the real-world performance of the proposed system. Our prototype system was demonstrated in the ACM S3 workshop and the Cloud Security Services (CLoSe) Workshop [31], [33].

### A. Securebox

The primary components of Securebox (SB) are an SDN controller used for enforcing network policies and gateway management, Open Virtual Switch (OVS) for network level functions and a policy-database for storing the network policies.

For the two prototype of Securebox, we used Fit-PC3 pro-Linux (fitPC) and Raspberry PI (R-Pi) [55], [54]. Table II gives a comparison of hardware specification of both devices. Section V gives a detailed evaluation of performance achieved by both version of Securebox. Our implementation of Securebox uses Floodlight SDN controller v1.1 at minimal configuration and OVS version 2.4.0 [52], [53]. A backup copy of Pol-DB (to be used in case of reboot) is also stored in the local file system. Policy table is currently implemented using hash tables but bloom filters can also be used [32].

**Portability**: Deploying Securebox on Raspberry Pi sized devices makes it much more portable for personal use. Users can carry Securebox and connect it to any available (insecure) Internet connection e.g. public Wi-Fi, hotel networks. Users then enable the option for setting up a secure personal access point (S-PAP) and connect their personal devices to the S-PAP. This approach will prevent any malware, spyware etc. on the insecure network from infecting user devices. It also prevents illegal access to user's devices connected to insecure network.

|  | Raspberry PI 2 (Model B) | Fit-PC3 pro Linux |
|---|---|---|
| CPU | 900Mhz Quad-Core | 1.6 Ghz Dual Core |
| Memory | 1 GB | 4 GB |
| Storage | SD Card | 320 GB |
| Ethernet | 1 | 5 |
| Wireless | None | 802.11 b/g/n |
| USB interface | 4 | 6 |
| HDMI | Yes | Yes |
| Cost | USD 35 | USD 533 |

Table II: Comparison between Raspberry-Pi and Fit-PC3

### B. Security Service

Security Service in the early prototype system was deployed using the OpenStack platform to dynamically deploy Docker containers running a simplified version of SNORT as an IDS instance and a Firewall service [42], [40], [41]. The choice of using Docker-based infrastructure instead of virtual machines was taken due to performance benefits of Docker containers [12]. However, that discussion is out of scope for this paper. Cloud Manager program was implemented to manage Docker containers, request handling of client events and load balancing across dockers. Cloud manager is also responsible of disseminating the client networks with the current network policies. Currently, we are also evaluating Kubernetes as a platform to deploy CSS [50].

## V. System Evaluation

We evaluate the prototype system against different scenarios. In order to minimize latency and maximize privacy for user, and as well as performance gain by using collaborative threat detection/mitigation mechanism, we have upgraded the system design multiple rounds.

### A. Latency

Latency is an important factor for user experience. As identified in [49] higher latencies can result in significant drop of website business. The proposed system design is susceptible to increase latency because the traffic analysis is done in the cloud environment.

In order to minimize the latency experienced by user, only metadata (by default: 6 tuple) information from the initial connection request is sent to CSS for analysis. Policies are cached locally to be used for any subsequent similar connection request. When a user accesses a website e.g. Youtube [2] for the first time, Securebox does not find a matching policy in Pol-DB and contacts security service

---
[2]www.youtube.com

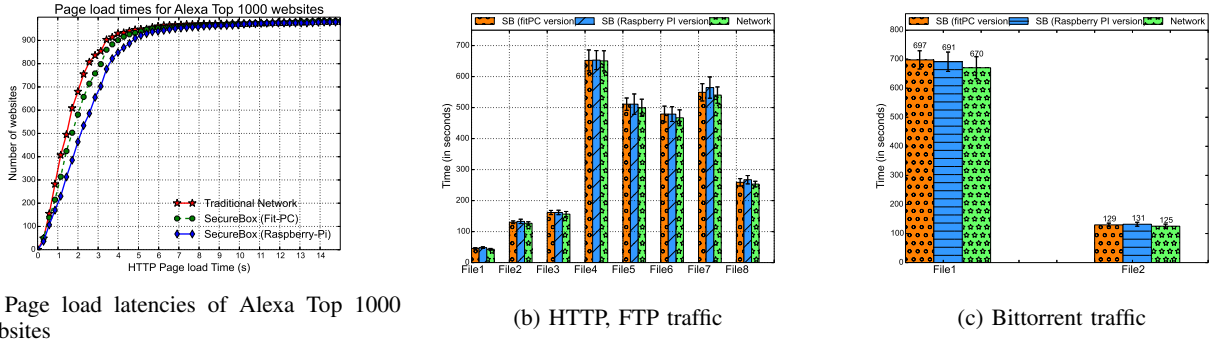

(a) Page load latencies of Alexa Top 1000 Websites

(b) HTTP, FTP traffic

(c) Bittorrent traffic

Figure 5: Performance comparison for user experienced latency

|  | Network | SB (fitPC) | SB (R-Pi) |
|---|---|---|---|
| Download Speed (Mbps) | 13.1 | 12.905 | 12.6 |
| Upload Speed (Mbps) | 2.153 | 1.783 | 1.69 |
| Download Consistency | 80% | 78% | 78% |
| Upload Consistency | 86% | 83% | 82% |
| Download BW (Mbps) | 18.504 | 17.693 | 17.296 |
| Jitter (server→ client) ms | 3.3 | 6.4 | 7.8 |
| Jitter (client→ server) ms | 5.8 | 7.6 | 8.2 |
| Packet loss (client→ server) | 0% | 0% | 0.50% |
| Packet loss (server→ client) | 0% | 0% | 0% |
| **MOS Score** | 4.3 | 4.1 | 4 |

Table III: VOIP Performance Comparison

to get a security policy for this connection request. This procedure will introduce a marginal increase in page load time of Youtube from 3.53 seconds to 3.81 seconds for the first time. Any subsequent requests will be addressed by matching policies in the in-memory Pol-DB and user will experience (almost) no latency.

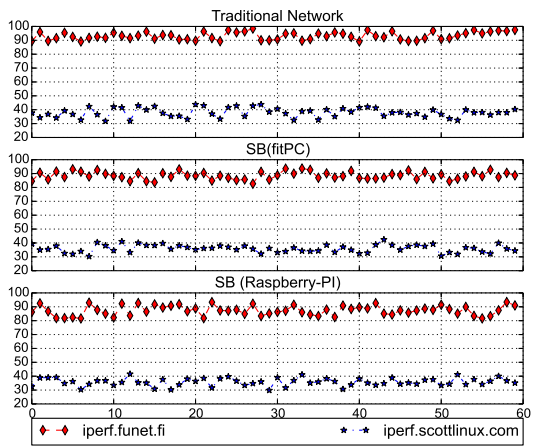

Figure 6: IPerf bandwidth testing

Figure 5a shows the CDF graph for page load times for Alexa Top 1000 websites (as of 23rd October 2015). It shows that Securebox does not introduce significant latency compared to the latency experienced with traditional networks. Figure 5a also shows that Raspberry Pi version introduces almost similar latency compared to the Fit-PC3, with much powerful hardware. The similarity in the performance is achieved by system design which requires Securebox to act as a policy enforcer and does not put any computational burden. It enables us to implement Securebox on a low-powered devices running decent hardware, save costs and improve portability without sacrificing the performance.

Figure 5b shows the results for file transfer over HTTP and FTP from public internet servers. The results show that proposed solution only increase download times by negligible percentage. Similar performance is achieved for Bittorrent traffic as shown in Fig. 5c. Bittorrent is an interesting use case because the connected peers are constantly being updated, new connections are made and old connections are dropped. Therefore, there are many traffic analysis requests being sent to CSS. However, the latency experienced by Securebox is almost similar to that of traditional network.

Table III shows the performance achieved for Voice-over-IP (VoIP) traffic. The jitter and consistency achieved by Securebox for both uplink/ downlink traffic is similar to that of a traditional network. Raspberry Pi version performs equally good as Fit-PC3 based version, achieving similar *mean opinion score* (MOS) score of 4, which shows that proposed system can deliver good quality unjittered VOIP traffic.

Pol-DB updates described in section III-C3 also contribute in minimizing the latency. They provide aggregated policy updates to the connected Secureboxes. It will increase the chances of finding a matching policy in the local Pol-DB of Securebox and reduces the number of requests made to CSS for decisions on new connection requests, hence minimizing the latency.

Figure 6 show the results obtained for bandwidth testing using iperf servers in the Internet. Results shows that Securebox (both fitPC and R-Pi version) achieve similar bandwidth as achieved in traditional networks.

*B. Collaborative Threat Detection/Mitigation*

Collaborative effort for detection of attacks and threats in the network is another key contribution of the proposed system.Section III-E explains how the Secureboxes installed in different network segments can act as network sensors which collect and send information to the CSS. The security service analyzes the information to detect attacks going on in disjoint networks. This broader view enables the Security Service to detect threats in various networks before they can substantially affect user in the network. Section III-C3 shows that when a new attack is detected in any network segment, Security Service dispatches Pol-DB updates to provide other networks with security policies to mitigate similar attacks.

Traditionally, attacks on different networks are detected long after they have infected the network. Once the attacks are detected, there is no mechanism for sharing this information with other entities because of legal and business reasons. Attackers exploit this lack of information sharing among network security teams and launch successful attacks on different organizations using similar techniques. Security professionals in each of these organizations face difficulties in detecting them and organizations have already suffered damage by the time these attacks are blocked.

Security community has long acknowledged the need for a mechanism for sharing network attack related information and there exists an IETF working group developing protocols for sharing information about network attacks [51].

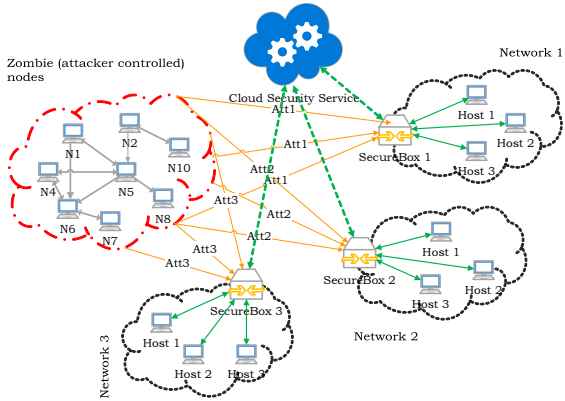

Figure 7: Attacker controlled nodes launching attack upon disjoint network segments

Figure 7 shows three disjoint network segments, each connected through a Securebox to CSS and the Internet. The attacker launches a port scanning attack on the hosts in these network segments [39]. Figure 8 shows the traffic analysis trends and the performance gain by using collaborative threat detection approach. A total of 20 nodes were used for generating traffic and a swarm of 15 zombie (attacker controlled) nodes each scanning 1000 random ports were used to attack three network segments each with three connected hosts

Figure 8a – 8d show the extracts from the attack traffic. These figures show the total attack traffic received, analyzed and dropped during the attack and compares the performance gain using collaborative approach.

Figure 8a shows that when the attacker initially launches an attack on network 1. Securebox 1 initially sends all the traffic to CSS for analysis. When CSS identifies the attacking nodes, it directs Securebox 1 to drop any more traffic coming from attacking nodes. Figure 8a shows that no traffic is dropped in the beginning. But, the volume of traffic dropped gradually increases as the CSS identifies more attacking nodes. The amount of traffic analyzed also decreases because Securebox 1 directly drops the traffic without sending it to CSS.

Following collaborative approach, CSS generates a Pol-DB to inform other network segments about the attacker nodes detected from network 1 activity. Figure 8b shows that Securebox 2 uses this information and drops all the traffic from suspicious nodes. It prevents any traffic from already detected attacker nodes to reach hosts inside the network, therefore minimizing the malicious traffic entering the network. Meanwhile, this approach also reduces the volume of traffic anlayzed by CSS.

CSS generates another Pol-DB update containing the list of attacker nodes detected by analyzing traffic from network 2 activity. Using this information, Securebox 3 is able to identify and drop ¿90% of the attack traffic without analyzing it. Figure 8c shows that volume of traffic analyzed is dropped to ¡10% of the total traffic received and all traffic from attack nodes is immediately dropped at network entry point.

With no collaboration in place, network 1, 2 and 3 will need to process all the traffic initially and traffic is dropped when CSS detects attacking nodes, as shown in Fig. 8d. No attack information is shared between the networks so every network needs to (re)detect the attacking nodes individually which allows some of the malicious traffic to penetrate the network before its detected.

*C. Privacy*

When network level statistics are shared with a third party service and user traffic is anlayzed in the cloud environment, privacy becomes an important concern. We recently conducted a detailed user study asking our respondents "*How comfortable would you be in sharing your network traffic statistics to get better network security and management*". Majority (¿ 60%) of the participants were comfortable with sharing their data with third party providing network management and security services. Recent research has explained that it is difficult to provide complete accountability while offering complete anonymity to the users [17]. Therefore, our proposed system only uses bare minimum information from user to find a suitable trade-off between user privacy and usable security.

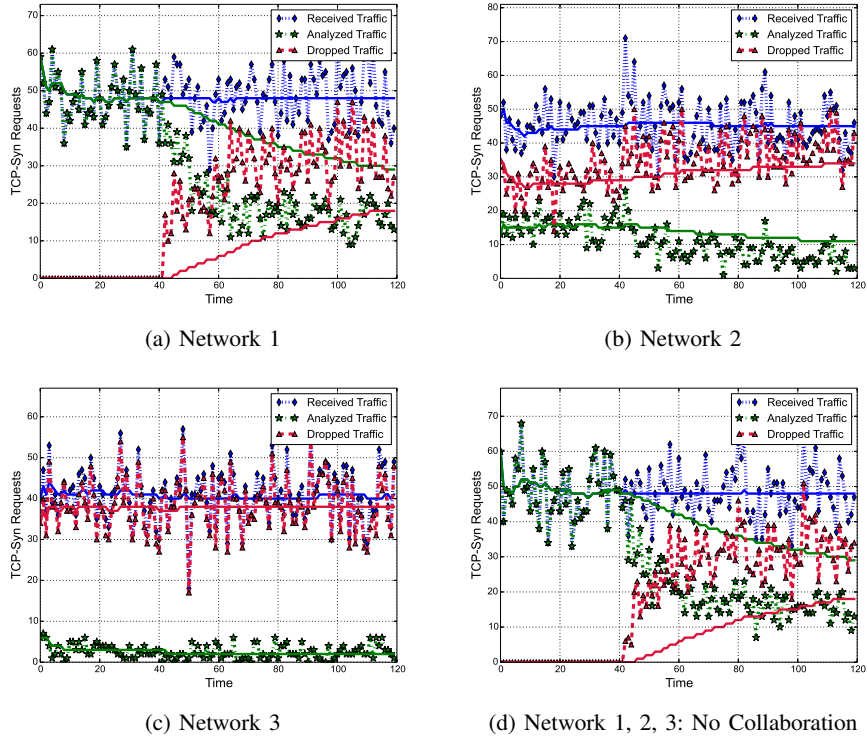

Figure 8: Performance gain in attack detection using collaborative approach compared to traditional approaches

When CSS is deployed by ISP, see section III-E, the privacy does not become an issue because ISP already performs a number of analysis on user traffic (to provide better QoS). In-house deployment of CSS alleviates all privacy concerns since it gives complete control over the infrastructure, data sharing and analysis. It can perform custom traffic analysis and manage Secureboxes deployed in the network with personalized security and management policies.

When CSS is deployed by a third party, different subscription models, see section III-D, allow user to decide what kind of services do they need for network management and how much data they want to share. Both, user and the service provider agree to the terms, which explicitly dictates how user data will be stored, analyzed and shared by the service provider. Users can opt-out to share their network traffic data being analyzed, by paying a subscription fee for the services and leased resources (i.e. middleboxes, traffic analysis). CSS provider can offer free management services or premium services to the users who agree to share their network traffic data for analysis purposes.

### D. Cost Efficiency

The cost of middleboxes range from hundreds to thousands of dollars a piece. Hiring network security professional's services for managing and updating these middleboxes can cost hundreds of dollars per hour. The update cycle of these middleboxes is a couple of years after which they need to be replaced by brand new equipment supporting more functionalities and higher bandwidth. A new set of equipment comes with deployment costs of their own. The high cost of deploying and maintaining network security is one of the reason why security is often neglected. Large enterprises can dedicate resources to manage these high costs but the problem is direr in the small enterprise and home networks.

In typical enterprise environment, middlebox deployment is underutilized most of the times and becomes a bottleneck during peak access periods. The proposed solution provides a cost efficient solution following *pay as you use* model. The proposed model deploys bare minimum number of middleboxes at all times, thus managing the resources more efficiently. During peak access periods, the deployment automatically scales to prevent middlexboxes from becoming the bottleneck for network traffic and reduce customer churn. SDN allows us to efficiently distribute traffic load among the middleboxes deployed in the cloud. In the proposed model, the cost of replacing a non-functional middlebox is much low compared to the cost of replacing a physical middlebox.

Table II shows the Securebox deployment using Raspberry PI like device costs ¡ USD 50 and provide almost similar performance as achieved by traditional networks, see section V. [24] also gives a detailed discussion on the how much cost is saved by deploying middleboxes in the cloud environment instead of deploying them physically at network vantage points.

### E. Scalability

Securebox design supports scalability due to its small size and minimal management overhead. User only needs to deploy the Secureboxes at network vantage points and connect them to CSS, which will ensure that all Secureboxes operate with consistent network policies. This model is also useful in enterprise environments where networking team only deploys Secureboxes at new establishments and offices and connects them to the CSS which installs initial set of policies and later makes sure that every Securebox has a coherent set of policies for operation. Network managers no longer need to individually configure middleboxes, switches or routers, saving resources and time. This exercise also minimizes any chances of inconsistent policies across network devices (which are very difficult to detect in the operational environment).

Section V-D explains the problems with traditional middleboxes e.g. underutilization, lack of scalability, update cycle and high costs etc. These problems incur high costs on already expensive network security infrastructure for an enterprise. Any network outages or bottlenecks can cause significant business losses. Therefore, the proposed system uses cloud resources to deploy traffic analysis services which can scale in real time while maintaining lower costs. Sherry et al. also supports our claim that using cloud resources to deploy middleboxes can reduces costs for enterprise security infrastructure [24].

### F. Attacks against system

The proposed system is susceptible to attacks by rogue Secureboxes and compromised CSS. Individual or set of rogue Secureboxes (working in collusion) can repeatedly generate malformed requests to CSS, compelling CSS to generate policies which mark suspicious/ insecure connections as safe. Rogue Secureboxes can also launch a DoS attack on the security service. On the other hand, a compromised security service can generate security services which force all Secureboxes to connect all traffic to any destinations. Any rogue node in the network can also claim to be a security service and start generating security updates to the Secureboxes in the network.

Figure 3 highlights the *Certification Authority*, which is responsible of managing and issuing certificates to connected Secureboxes. When a Securebox is registered, a certificate is generated for it to communicate with CSS. All updates from CSS are encrypted and signed. Securebox will reject any updates coming from sources other than the registered CSS and report those sources back to the CSS. Similarly, CSS logs the traffic analysis request and any Securebox showing suspicious behaviour i.e. generating fake or repeated requests, is blacklisted. The proposed system design does not allow Secureboxes to communicate with each other, which prevents any chances of Securebox generating false Pol-DB updates to each other.

### G. Uplink bandwidth saving

User bandwidth is a precious resources and the proposed system ensures that security service has minimal impact on the user bandwidth. Every traffic analysis request for a new connection contains ¡40 bytes of information. The Pol-DB updates also help in minimizing the number of traffic analysis request generated by the user. Section III-C3 explains that scheme followed by Pol-DB updates which uses periods of less traffic activity for generating updates.

## VI. RELATED WORK

Following the initial proposal of ETHANE to manage control plane in runtime, researchers showed the possibilities of how this technology could revolutionize the traditional networking [38]. SDN has been a hot topic in academic research community since 2010. Google's announcement for using SDN to control its inter-datacenter traffic routing further increased the popularity of SDN with increased deployments in real-time traffic [56]. Both the academic and industrial community proposed a number of techniques for improving SDN performance, fault tolerance. However, SDN has been mainly deployed in large enterprise to manage WAN and data center traffic [37]. This paper is the first attempt to exploit the potential of SDN for managing SOHO and SME networks including IoT and BYOD environments.

### A. Academic Research

SDN research has mainly focused on improving performance and fault tolerance of SDN controllers and better communication with OVS hardware. Different SDN controllers have been proposed by researchers offering better security, resilience to attacks. *Elasticon* is proposed in [36], which is a distributed SDN controller to minimize impacts of DDoS attacks against the SDN controller. Besides improving the security for SDN, researchers have also been investigating how to apply SDN to enhance security for mobile and wireless networks [28], [29].

SDN and network function virtualization (NFV) has opened new possibilities to improve network management. Researchers have also explored the possibility of virtualizing the middleboxes to achieve reduced costs and increased scalability. Sherry et al. have proposed to deploy virtual middleboxes in the cloud environment and showed that there is no significant performance degradation [24]. A model offering DPI as a service has been proposed in [11]. Their work complements our proposed system for setting up the virtual middleboxes in the CSS. Deidtect proposes the use of SDN to dynamically re-route network traffic through a centralized middlebox deployment (possibly in the cloud environment) [27]. Their solution only explores SDN support for dynamic re-routing of traffic in enterprise environment but does not provide significant evaluation of its performance. Our work exploits SDN's potential for managing network policies, monitoring per device communications, interactions and dynamic re-routing of traffic through CSS. The proposed CSS offers many more features compared to a standard middlebox deployment.

SENSS proposes an interface design for ISPs which can provide user's traffic statistics on demand [26]. Users can then make an analysis if they are under an attack and request ISP to take user-directed actions to secure their networks. This proposal requires significant user interaction to analyze traffic traces and make decision about what should be done to secure the network. ISPs might not be comfortable to provide an interface to users for accessing traffic statistics and manipulating traffic propagation. Alwabel et al. proposes a model where users can classify their traffic from the ISP and decide what path should it follow to their network e.g. users would want class A traffic to be redirected through middleboxes, whereas class B traffic coming directly to their network [22]. Once again, ISP may not allow the users to control the routing of their traffic.

Researchers have proposed a system to control the wireless AP using SDN [35] but their work is focused on AP management and does not focus on the security aspect. This work can be used by Securebox deployments in wireless environments. *Resonance* has been proposed for securing enterprise environments by providing dynamic access control on flow level information [13].

Recent set of events have greatly increased public awareness of privacy of their data and people are concerned for the privacy of their data and devices more than ever before. Hacking, online scams, digital extortion and state sponsored cyber espionage activities has become an increasing concern for the users. Many contemporary devices contain sensors collecting information about the environment and the users [43]. New generation of smartphones contains tens of these sensors and there have been plenty of cases where smartphone and mobile devices have been used to spy on the user [62].

Researchers have been working on the idea of how to make smartphones more secure so that they do not leak sensitive information about the user to unwanted entities. *Databox* has been proposed in [30], which collects all the information about the user and shares it only with authenticated third parties. OpenPDS based service provides "*safe answers*" to the third parties asking for data collected from the users [8]. These "*safe answers*" are designed not to leak any related information about the users.

Securebox can offer similar functionalities for IoT and mobile devices. An IoT hub module in the Securebox can collect data from the IoT devices connected to the network and provide an interface for third party applications to access this data. Securebox will provide authenticated and validated access to the collected information from user specified applications. Securebox will thoroughly audit each incoming data request and only respond to queries coming from trusted sources. However, this functionality will require support from industrial manufacturers and third party services to work with this architecture.

*B. Industrial Research*

Many industrial actors have been trying to improve gateways to make them smarter and automatically manageable. Recently, Google announced OnHub solution for home environments [57]. OnHub currently manages home wireless network environment and provides automatic management services via Google-On application but Google OnHub comes at a heavy price tag of USD 219. Qualcomm $^{\circledR}$ has presented a secure home gateway and IoT hub by combining Gigabit Wi-Fi support of Qualcomm$^{\circledR}$ VIVE 802.11 and Qualcomm$^{\circledR}$ StreamBoost technology [58]. It provides security features like parental control and automatic virus detection features. Qualcomm has included a high performance processor so that the device can learn user actions and mimic them later.

Home gateway initiative (HGI) is an alliance of leading home gateway manufacturers working together to improve the home gateway experience [59]. HGI's role is to specify requirements and plans for home gateways that support QoS and roll-out of triple play and broadband services. HGI work on enabling services to include delivery framework for smarthome services. ProSyst series of products launched by Bosch Group provide multi service platform for smart homes. They provide a framework for developing smart home applications and creates a market place for it to benefit third party developers and open source standard platform [60].

Currently available IoT hubs do not offer many features and are difficult to operate but a number of manufacturers have been developing new technologies for smarter home gateways and IoT hubs by including features like automated management. Most of this work has been targeted to get better wireless coverage, easier device communication, higher bandwidth available for the connected devices. However, security aspect is not considered in most of the cases. Our proposed system bundles the management features and access point control along with better security functionalities. It also provides features like parental control, restricted D2D communication, better traffic analysis and middlebox functionalities etc.

## VII. Conclusion

Our proposed system is the first realization of the idea to bring remote management, enterprise grade security and benefits of SDN to SOHO and small enterprise networks. Securebox is an easy to deploy, highly portable solution with much lower cost. System evaluation has shown that it can be deployed on Raspberry Pi sized devices which increase portability and reduce costs. CSS offers subscribers a facility to analyze their traffic through middleboxes in cloud environment. It also enables remote management of Secureboxes that are installed at network edge, to take the burden of network management away from users. Offloading the traffic analysis and network management operations to CSS increases the system scalability for home and enterprise networks. Secureboxes installed in different networks act as a sensor for CSS and collect network level statistics from these networks. CSS can use this data to perform analysis to detect botnets, malwares and other insights about the network. The result of their analysis can be used to detect, threats to network security, improve QoS and management for subscriber's networks. Our proposed system introduces a collaborative scheme which allows networks to share attack related information which helps in rapid detection and mitigation of attacks on disjoint network segments.

Our work shows that using SDN in SOHO and small enterprise environments does not degrade user experience. The system is designed to minimize impacts on network latency and user privacy. Our experiment results show that Securebox introduces only a marginal (almost negligible) increase in latency experienced by the user. We also introduce a number of subscription models for CSS to ensure user privacy. CSS analyzes the data collected from connected networks to detect various abnormal behaviors and threats occurring in those networks. The collaborative scheme introduced in this paper allows CSS to rapidly share this information with all Secureboxes so that they can promptly block any attempts to attack the network.

Our prototype system has shown that the proposed model for securing home and enterprise environments is functional and effective. It can be deployed incrementally with current infrastructure and can resolve many security and network management problems in traditional networks. It can ease the tasks of enterprise network management teams. The proposed model also resolves many security issues encountered in IoT and BYOD environments.

## Acknowledgment

This work is carried out in the DIGILE Internet of Things (IoT) and Cloud Security Services (CloSe) projects supported by Tekes and Academy of Finland. The authors would like to thank Valtteri Niemi, Seppo Hätönen, and Emad Nikkhouy for their feedback and technical support.